\documentclass{article}

    \usepackage[preprint]{neurips_2025}

\usepackage[utf8]{inputenc} 
\usepackage[T1]{fontenc}    
\usepackage{hyperref}       
\usepackage{url}            
\usepackage{booktabs}       
\usepackage{amsfonts}       
\usepackage{nicefrac}       
\usepackage{microtype}      
\usepackage{xcolor}         
\usepackage{graphicx}
\usepackage{subcaption}
\usepackage{float}
\usepackage{amsmath}

\title{The Wisdom of Agent Crowds: A Human-AI Interaction Innovation Ignition Framework}

\author{
Senhao Yang\\
University of Chinese Academy of Sciences\\
\texttt{yangsenhao24@mails.ucas.ac.cn}
\AND
Qiwen Cheng\\
Faculty of Psychology\\
Beijing Normal University\\
\texttt{psychengqiwen@163.com}
\AND
Ruiqi Ma\thanks{Corresponding author}\\
Sino-Danish College\\
University of Chinese Academy of Sciences\\
\texttt{maruiqi23@mails.ucas.ac.cn}
\AND
Liangzhe Zhao\\
College of Computer Science and Technology\\
Fudan University\\
\texttt{22302010032@m.fudan.edu.cn}
\AND
Zhenying Wu\\
School of Business\\
China University of Political Science and Law\\
\texttt{wzyscience@163.com}
\AND
Guangqiang Yu\\
University of Chinese Academy of Sciences\\
\texttt{20210608208@smail.lnu.edu.cn}
}

\begin{document}

\maketitle

\begin{abstract}
With the widespread application of large AI models in various fields, the automation level of multi-agent systems has been continuously improved. However, in high-risk decision-making scenarios such as healthcare and finance, human participation and the alignment of intelligent systems with human intentions remain crucial. This paper focuses on the financial scenario and constructs a multi-agent brainstorming framework based on the BDI theory. A human-computer collaborative multi-agent financial analysis process is built using Streamlit. The system plans tasks according to user intentions, reduces users' cognitive load through real-time updated structured text summaries and the interactive Cothinker module, and reasonably integrates general and reasoning large models to enhance the ability to handle complex problems. By designing a quantitative analysis algorithm for the sentiment tendency of interview content based on LLMs and a method for evaluating the diversity of ideas generated by LLMs in brainstorming based on k-means clustering and information entropy, the system is comprehensively evaluated. The results of human factors testing show that the system performs well in terms of usability and user experience. Although there is still room for improvement, it can effectively support users in completing complex financial tasks. The research shows that the system significantly improves the efficiency of human-computer interaction and the quality of decision-making in financial decision-making scenarios, providing a new direction for the development of related fields.
The available github code is at \href{https://anonymous.4open.science/r/The-Wisdom-of-Agent-Crowds-A-Human-AI-Interaction-Innovation-Ignition-Framework-1FF5}{Anonymous Github}.

\end{abstract}

\section{Introduction}
Human-computer interaction and collaboration represent a critical step in forming multiple dimensions of thinking, thereby igniting innovative and creative vitality. Brainstorming, as an effective method, generates valuable and novel ideas through cognitive collisions formed by interpersonal interactions. Since the release of ChatGPT, large language models (LLMs) have demonstrated capabilities comparable to humans in various aspects of natural language processing. Currently, LLM-based agents and workflows are thriving. Employing LLMs to simulate experts with diverse backgrounds in brainstorming—leveraging their broad knowledge and generalization abilities to solve domain-specific problems through interdisciplinary thinking or provide novel perspectives—warrants in-depth exploration.  

However, objective evaluation metrics for non-QA data remain lacking. Existing approaches typically rely on "LLM-as-a-judge" or human scoring, introducing significant uncertainty and subjectivity. In this study, we develop a brainwrite-based human-computer interaction framework aimed at integrating human supervision and cross-lingual multi-agent discussions to stimulate collective wisdom.  

Prior research has confirmed that multi-agent collaboration significantly enhances the quality and efficiency of information generation. For example, \cite{wang2024autopatent}  proposed AutoPatent, an automated patent-drafting framework utilizing eight agents across three categories to complete full patent writing. Further, multi-agent collaboration improves model performance and robustness: \cite{wei2025dontjustdemoteach} introduced a principle-based multi-agent prompting strategy, showing that multi-agent systems can compensate for the limitations of single models through division of labor and mutual supplementation. \cite{wang2025lns2} combined LNS2 with multi-agent reinforcement learning (MARL) to propose LNS2+RL, significantly enhancing success rates in high-density path planning.  

Despite above potential, multi-agent collaboration faces challenges, particularly the trade-off between security and collaboration efficiency. \cite{peignelefebvre2025multiagentsecuritytaxtrading} highlighted that malicious instructions can propagate through multi-hop networks, affecting the entire system—mitigating attack risks but compromising collaboration efficiency. Additionally, larger multi-agent systems increase coordination and computational complexity, making reasonable architectural design crucial. \cite{taguelmimt2025multiagent} developed the SALDAE algorithm for coalition structure generation (CSG), enabling rapid discovery of high-quality solutions in large-scale problems involving thousands of agents.

In this work, we construct a multi-agent brainstorming framework rooted in the Belief-Desire-Intention (BDI) theory, which plans tasks according to user intent, reduces cognitive load through real-time structured text summaries and an interactive Cothinker module, and integrates general and reasoning large models to enhance complex problem-solving capabilities. We designed a method to evaluate idea diversity using k-means clustering and information entropy, develop an interactive system (X), and conduct extensive human factors experiments, surveys, and interviews to assess human-AI collaboration effects. User sentiment analysis provides insights into improving satisfaction.

Our research contributions are as follows: (1) We adopt a human-computer interaction approach based on the Belief-Desire-Intention (BDI) theory and implement an intention recognition function centered on user needs.  (2) The system effectively reduces users’ cognitive load in understanding and operating complex systems by providing real-time updated structured text summaries and interactive thinking aids.  (3) We stimulate collective wisdom through multi-expert discussions and human supervision.  

Our findings include:  \\
•	Brainwrite significantly improves diversity metrics, and Chain-of-Thought (CoT) prompting outperforms zero-shot prompting. Entropy values of large language models (LLMs) differ significantly across different language environments.  \\
•	Human factors testing shows that System X has moderate usability, effectively supporting users in completing complex financial tasks, but there is room for improvement in interactive experience.  \\
•	User sentiment toward System X is predominantly positive, though aspects such as personalization and functional completeness require further enhancement.

\section{Related Work}
\subsection{The wisdom of multi-agent collaboration System}

Research on multi-agent collaboration based on large language models (LLMs) is relatively mature. Existing literature has explored domains such as debate and text classification \citep{wei2025dontjustdemoteach}, path planning \citep{zhou2024looselysynchronizedrulebasedplanning}, coalition structure generation \citep{taguelmimt2025multiagent}, image generation \citep{xie2025anywheremultiagentframeworkuserguided}, and urban revitalization \citep{ni2024planninglivingjudgingmultiagent}. These studies demonstrate that multi-agent systems can effectively aggregate information \citep{huang2025structuredreasoningfairnessmultiagent} and improve task processing efficiency and accuracy through division of labor—for example, in text classification tasks \citep{wei2025dontjustdemoteach}. Additionally, collaboration among agents enhances system robustness and adaptability, such as maintaining effective cooperation under communication delays \citep{song2025codecommunicationdelaytolerantmultiagent} and enabling collision-free path planning in high-agent-density environments. The role of multi-agent collaboration extends beyond performance improvement; it also promotes exploratory behavior and diversity \citep{chen2024noveltyguideddatareuseefficient}, enhances sample utilization efficiency, and facilitates solving complex knowledge-intensive tasks \citep{yue2025synergisticmultiagentframeworktrajectory}. For instance, in path planning, the LNS2+RL algorithm demonstrates how integrating the advantages of different techniques can optimize both solution quality and computational efficiency \citep{wang2025lns2}.  

However, multi-agent collaboration still faces challenges such as information inconsistency and bias \citep{huang2025structuredreasoningfairnessmultiagent}, security and collaboration efficiency issues \citep{peignelefebvre2025multiagentsecuritytaxtrading}, and problems arising from asynchronous communication \citep{song2025codecommunicationdelaytolerantmultiagent}. Some scholars have attempted to introduce human supervision to enhance system reliability and fairness, aligning with human preferences. Our work aims to strengthen preference alignment in multi-agent generated content by incorporating human intention recognition, thereby stimulating collective intelligence.

\subsection{Cognitive model and Human-AI Interaction}

From a cognitive mechanism perspective, \cite{hitch1976verbal}'s  working memory model posits that human working memory has limited capacity. Cognitive Load Theory \citep{sweller1988cognitive} further indicates that when information input exceeds the processing capacity of working memory, it leads to a surge in extrinsic cognitive load, thereby reducing decision-making efficiency. \cite{nesbit2006learning}  noted in their research that visual tools such as mind maps can reduce the immediate burden on working memory by reorganizing complex financial information, thus decreasing extrinsic cognitive load. \cite{hollender2010integrating} integrated Cognitive Load Theory (CLT) with Human-Computer Interaction (HCI) to improve e-learning environment design, emphasizing the reference value of cognitive load concepts for HCI design principles. \cite{kosch2023survey} and \cite{darejeh2024exploring} proposed multiple methods for measuring cognitive load in HCI research. \cite{cui2025developing} developed the Chinese Human-Automation Trust Scale (C-HATS), which is significant for evaluating human-machine trust in various automated systems within the Chinese cultural context. Our study, based on practical case studies, investigates user satisfaction from the perspective of cognitive load to enhance interactive experiences.

\section{Preliminaries}
\subsection*{Brainwrite evaluation design based on LLMs}
Brainwrite is a creativity generation and problem-solving method where participants first independently record ideas, then stimulate innovative thinking through sharing and iterative development (similar to passing notes). This approach combines individual reflection with collective wisdom, mitigating the bandwagon effect of traditional brainstorming to promote diverse idea expression. When applied to LLM-based brainstorming, it controls incremental text length in each round, ensuring the model maintains focus on historical information.

\subsection*{Text Embeddings}
Text Embeddings is a technology that converts natural language into fixed-length numerical vectors, capturing textual semantics through these vectors such that closer vector distances indicate greater semantic similarity. Applicable at the word, sentence, and document levels, this technique is widely used in NLP tasks such as text classification and sentiment analysis. This paper employs the gte-Qwen2-7B-instruct model to transform input text into 3584-dimensional vectors for semantic representation.

\subsection*{K-means clustering}
For two Text Embeddings represented by vectors $\vec{A}$ and $\vec{B}$, their cosine similarity is defined as:

\[
\text{Cosine Similarity}(\vec{A}, \vec{B}) = \frac{\vec{A} \cdot \vec{B}}{||\vec{A}|| \ ||\vec{B}||}
\]

Where $\vec{A} \cdot \vec{B}$ represents the dot product of vectors $\vec{A}$ and $\vec{B}$, and $||\vec{A}||$ and $||\vec{B}||$ represent the magnitudes (or norms) of vectors $\vec{A}$ and $\vec{B}$ respectively. Cosine distance is the complement of cosine similarity and can be expressed as:

The silhouette coefficient $s(i)$ for a single sample $i$ has a value range of $[-1, 1]$ and its calculation formula is as follows:

\[
s(i) = \frac{b(i) - a(i)}{\max\{a(i), b(i)\}}
\]

For the silhouette coefficient of the entire dataset, it is usually the average of the silhouette coefficients of all samples, i.e.:

\[
S = \frac{1}{N} \sum_{i=1}^{N} s(i)
\]

Here, $N$ is the total number of samples in the dataset. The closer the silhouette coefficient $S$ of the entire dataset is to 1, the better the clustering effect; conversely, the closer it is to -1, the worse the clustering effect.

\subsection*{Entropy}
For a discrete random variable $X$, which can take values $x_1, x_2, \ldots, x_n$ with corresponding probabilities $P(x_1), P(x_2), \ldots, P(x_n)$, then the information entropy $H(X)$ of $X$ is defined as:

\[
H(X) = - \sum_{i=1}^{n} P(x_i) \log_b P(x_i)
\]

\subsection*{System design}

\begin{figure}[H] 
  \centering 
  \includegraphics[width=0.7\linewidth]{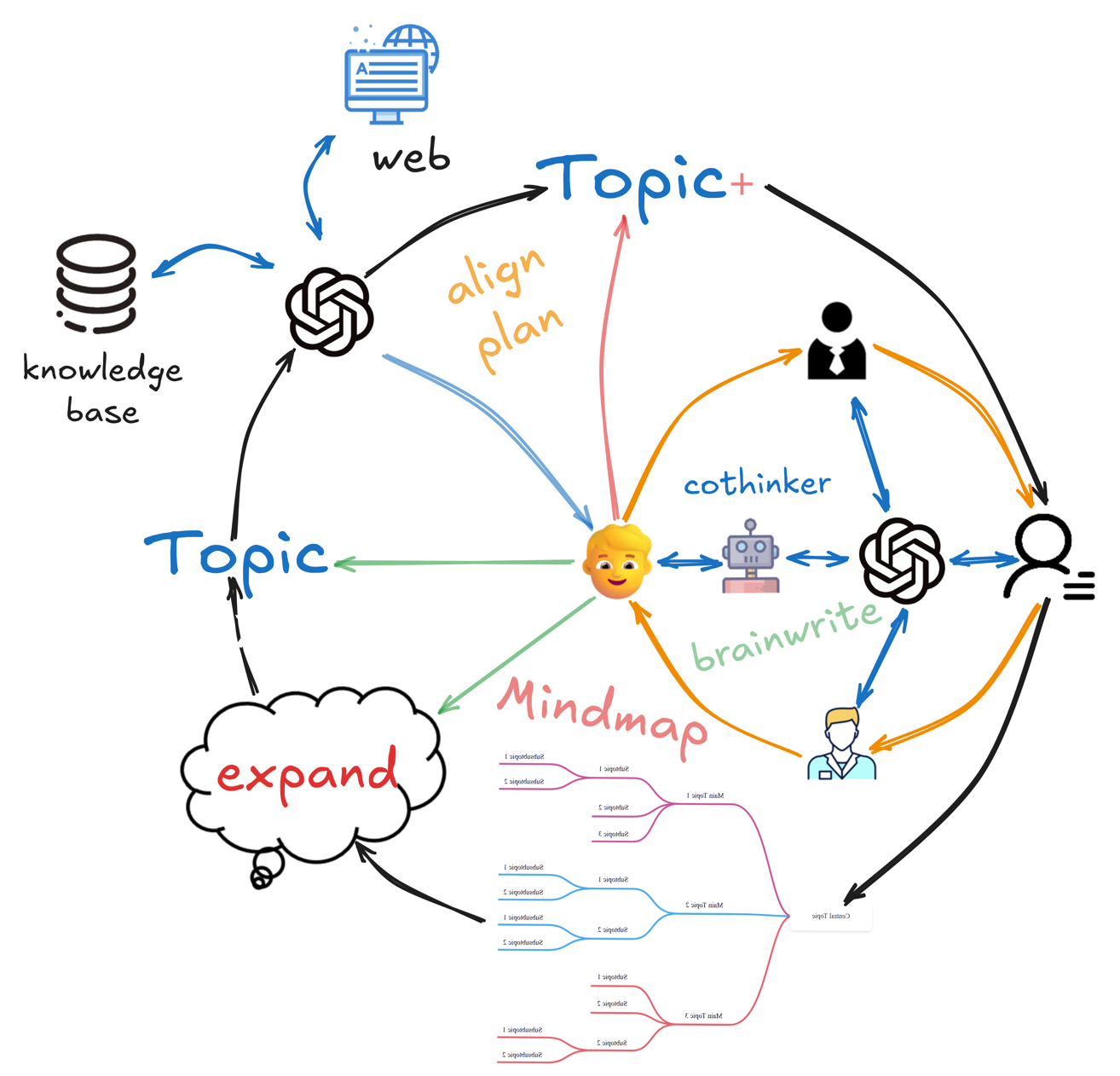}
  \caption{Brainwrite workflow} 
  \label{fig:Brainwrite workflow} 
\end{figure}

Fig\ref{fig:Brainwrite workflow} illustrates the three-part architecture of Brainwrite:

Human-Driven + LLM Assistance for Topic Definition and Preliminary Analysis: Users input financial topic keywords, and the LLM retrieves information from the internet and knowledge base to generate preliminary background analysis. Building on this, Cothinker provides intuitive (System-1) creative directions or thinking pathways to help users quickly engage in the creative process. Users review the content from LLM and Cothinker, refining it through additions, deletions, or modifications to ensure alignment with their needs and topic relevance.

LLM-Driven + Cothinker-Supported Deep Exploration and Thought Expansion: The LLM conducts multi-dimensional in-depth analysis (concepts, cases, trends), after which Cothinker supplements with deliberate (System-2) insights. Users synthesize information through interactive discussions, forming unique perspectives by integrating LLM outputs and iterative feedback.

User-Driven + Cothinker-Refined Output Generation and Optimization: Users draft initial deliverables (reports, plans, etc.) based on prior analysis. Cothinker provides optimization suggestions for language, logic, and innovation, leading to iterative improvements. The final outputs are refined through this process and applied for practical use.

\section{Experiments}
LLM brainstorming data collection is implemented using Zhipu company's API, and text data embedding is implemented using gte-Qwen2-7B-instruct deployed on NVIDIA A800 80G.
\subsection*{Brainwrite performance}
We generated 30 open-ended discussion topics for each of the "mathematical sciences," "finance," and "philosophy" domains using LLMs, and collected text data from brainstorming sessions under different models, workflows, prompting methods, and numbers of Agent participants.

This paper will divide the collected data under different conditions into 3 groups: $Group = G_1, G_2, G_3$. Each $G_x$ includes four classes: $bk, raw, spbk, spraw$. Each class within a group contains content generated using 3 models ($flash, air, plus$). Here, $bk$ indicates brainstorming under the condition of providing background prompt words; $raw$ indicates brainstorming without providing background prompt words; $sp$ represents a non-brainstorming format. $G_1$ adopts a zero-shot prompting strategy, $G_2$ adopts the CoT (Chain-of-Thought) method. $G_3$, under the zero-shot condition, sets different numbers of participants $C$, aiming to explore the impact of different Agent numbers on the diversity of generated viewpoints in LLM brainstorming.

\begin{figure}[H] 
  \centering 
  \includegraphics[width=0.7\linewidth]{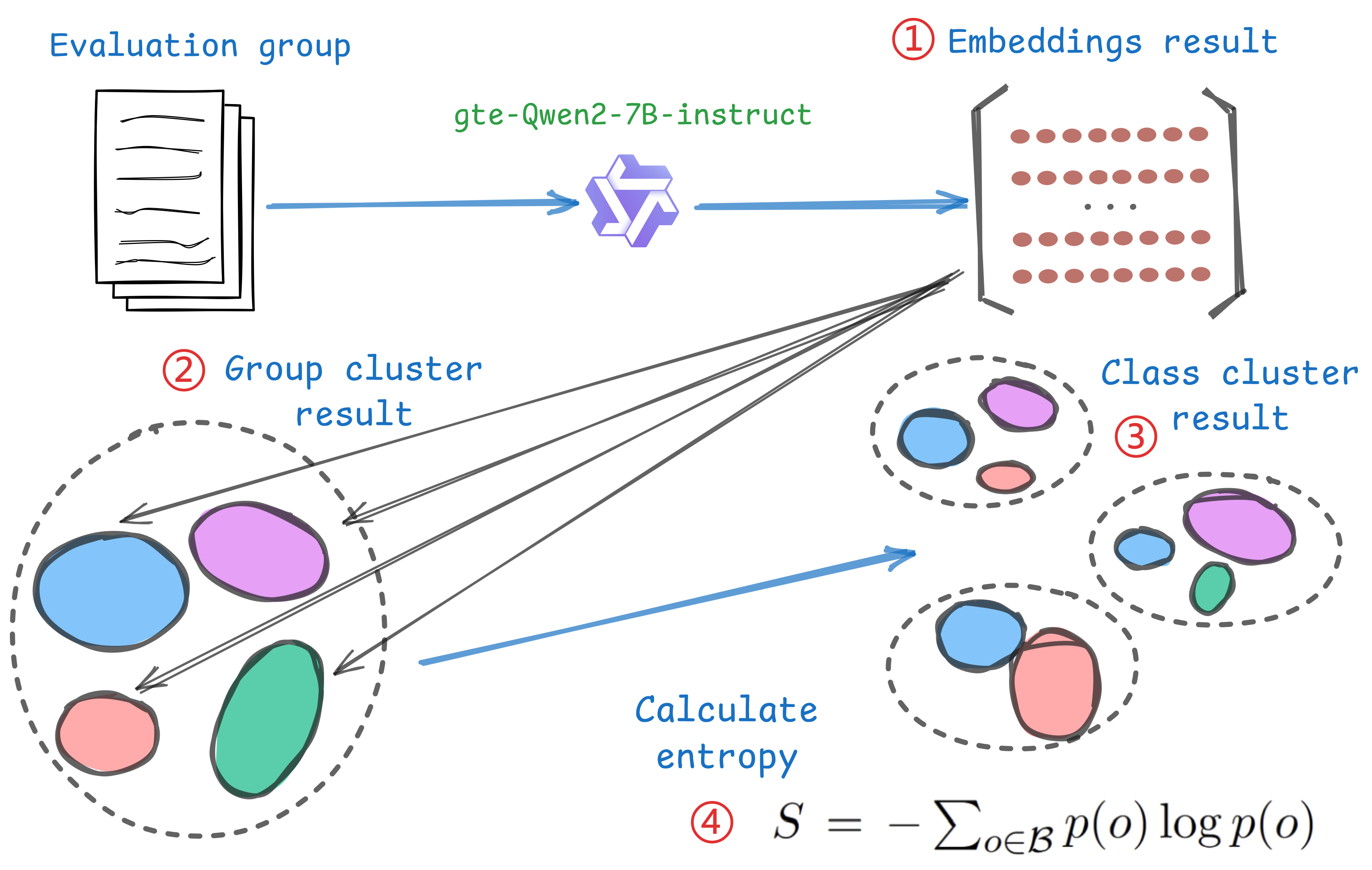}
  \caption{Topic diversity evaluation based on K-means-Entropy} 
  \label{fig:Topic diversity evaluation based on K-means-Entropy} 
\end{figure}

The collected textual data is vectorized using gte-Qwen2-7B-instruct, and clustering is performed based on the vectorized results for each $G$. During the clustering process, $k$ is chosen from 3 to 12, and the top 5 cluster results with the highest silhouette coefficient are saved. Subsequently, the saved cluster results of $G$ are mapped back to each $G$. Within each $G$, the probability of different occurrences of cluster results within $G$ is calculated, and information entropy is then derived based on these probabilities. As Figure \ref{fig:Topic diversity evaluation based on K-means-Entropy} shows.

\begin{figure}[H]
  \centering
  \begin{subfigure}[b]{0.45\textwidth} 
    \includegraphics[width=\textwidth]{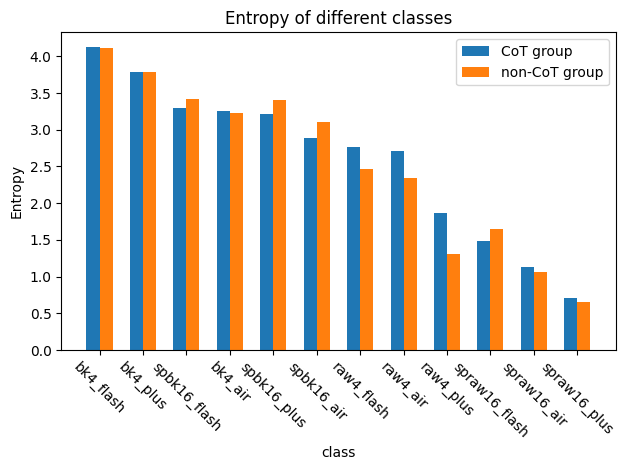} 
    \caption{Differences between groups and with/without Chain of Thought} 
    \label{fig:sub1} 
  \end{subfigure}
  \hfill 
  \begin{subfigure}[b]{0.5\textwidth}
    \includegraphics[width=\textwidth]{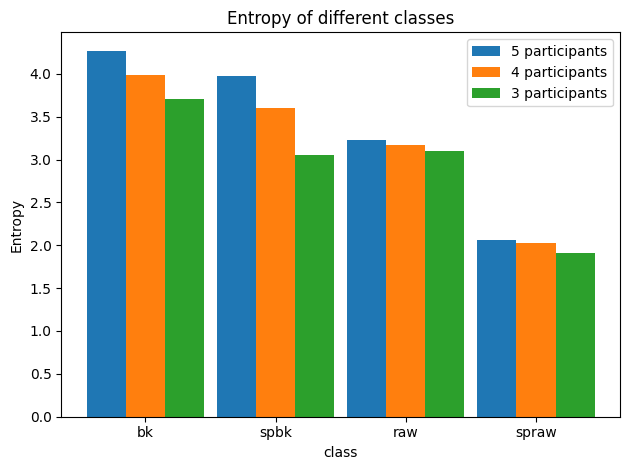} 
    \caption{Differences in the number of participants} 
    \label{fig:sub2} 
  \end{subfigure}
  \caption{Figure 3 Ablation study}
  \label{fig:ablation result}
\end{figure}

At the same time, our ablation experiments were conducted by setting different groups and numbers of participants. Figure 4 shows the results, indicating that when using Brainwrite and providing background prompt words to LLM experts, the scores of the evaluation metrics improved by 178.4\% compared to when LLMs generated opinions independently, and the use of the CoT prompting method was superior to zero-shot ($p < 0.05$). As Figure \ref{fig:ablation result} shows.

\subsection*{Human factor testing based on case study}
\begin{figure}[H] 
  \centering 
  \includegraphics[width=0.95\linewidth]{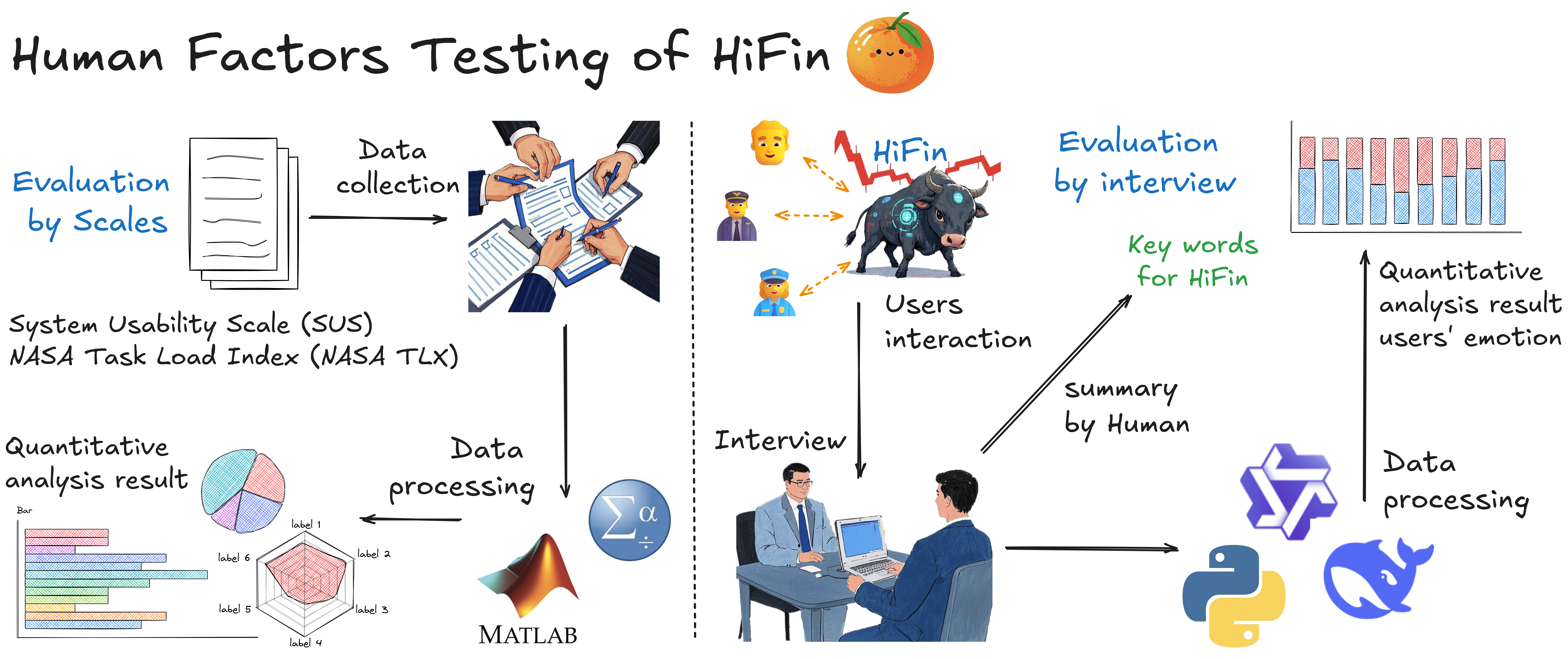}
  \caption{Human factor testing design} 
  \label{fig:Human factor testing design} 
\end{figure}

In the financial investment context, we developed a domain-specific interactive system (X) for case analysis and conducted human factors testing using two methodologies. First, two standardized scales were employed: the System Usability Scale (SUS) and the NASA Task Load Index (NASA-TLX). Second, user interviews were conducted following the framework illustrated below. The SUS was used to quantify system usability, providing numerical feedback to guide design optimization. The NASA-TLX was adapted for X's financial scenario by retaining its core dimensions: Mental, Physical, Temporal, Effort, Overall, and Frustration—to measure psychological and operational workload in high-risk decision-making, enabling cross-domain load comparisons through consistent dimensional definitions. 
As Figure \ref{fig:Human factor testing design} shows.

As of March 19, 2025, data collection yielded 118 questionnaires (58 SUS, 60 NASA-TLX) and 10 interview records. Scale data were processed using scientific computing tools, while interview data underwent manual summarization and quantitative sentiment analysis via large language model (LLM)-based next-token prediction.

\subsection*{User sentiment analysis}

We conducted sentiment analysis on interview data from system (X) human factors experiments. Interviews were semi-structured \citep{brinkmann201414}, conducted via one-on-one or group online meetings/face-to-face talks to flexibly capture participants’ views \citep{roulston2018qualitative}. With consent, we recorded feedback after participants used X system or watched demos. Traditional Natural Language Processing (NLP) methods like Latent Dirichlet Allocation (LDA) have limitations: LDA lacks direct sentiment calculation, while semantic modeling struggles to capture deep meanings and may miss key information. Thus, we used large language models’ next-token prediction, fine-tuning them to model interview text semantics and predict sentiment.

Using grounded theory \citep{charmaz2006constructing}, we performed qualitative analysis on interview data. Coding and theme extraction revealed multi-dimensional user feedback. Top frequent themes included "user experience," "usability," and "personalization," showing strong focus on operational convenience and customized services. For example, a user said, "Simple to use, but I hope it fits my habits better," affirming basic usability while expressing needs for personalized settings—indicating a need to optimize interaction flows for user habits.  

"System stability," "interaction efficiency," and "response speed" also mattered, reflecting concerns about technical performance. Feedback like "The system is too slow sometimes, affecting my experience" directly pointed to lag/delay issues, highlighting that technical stability/response speed—foundational to human-computer interaction—directly impacts user sentiment. Technical optimizations are needed to reduce frustration.  

"Privacy protection" and "data security" stood out as key themes, reflecting widespread data security concerns. A user noted, "The system is smart, but I worry about data leaks," indicating that while providing intelligent services, X system must strengthen data security to build user trust.  

Lower-frequency keywords like "interface design," "feedback mechanisms," and "learning curve" showed attention to experiential details. Suggestions like "Make the interface more attractive" or "Feedback isn’t timely enough" highlight that while not core pain points, these details cumulatively impact satisfaction and deserve attention in future improvements.  

Overall, analysis revealed users’ multi-dimensional needs for X system in usability, performance stability, data security, and detailed experiences—providing concrete directions for improving interaction flows, upgrading technical architecture, and formulating security strategies.

\section{Conclusion and Limitation}
In this study, we aim to construct an efficient human-machine collaboration framework for financial scenarios. By integrating human supervision with multi-agent collaboration, we seek to leverage collective intelligence and enhance the quality of complex decision-making. We have developed a multi-agent brainstorming framework based on the Belief-Desire-Intention (BDI) theory and an interactive system X, comprising three core modules: "Topic Definition and Preliminary Analysis (Human-led + LLM-assisted)," "In-depth Exploration and Thought Expansion (LLM-led + Cothinker-assisted)," and "Output Generation and Optimization (User-led + Cothinker-optimized)." This framework achieves user intent-driven task planning, reduced cognitive load, and enhanced viewpoint diversity.

Experiments and evaluations demonstrate that the proposed Brainwrite method significantly improves the diversity of generated viewpoints, and the Chain-of-Thought (CoT) prompting approach outperforms zero-shot prompting. System X exhibits moderate usability in financial tasks, effectively supporting users in completing complex decisions. User sentiment is predominantly positive, although there is room for optimization in personalized interaction and feature completeness. Quantitative evaluation based on k-means clustering and information entropy, along with qualitative analysis through user interviews, provides multi-dimensional improvement paths for system iteration.

In future research, we plan to explore multi-modal interaction features to enhance user experience and extend the framework to high-risk decision-making scenarios in other domains such as healthcare and education. Simultaneously, we will further optimize the multi-agent coordination mechanism and sentiment analysis model to improve the system's dynamic adaptation to user intentions.

Our research has certain limitations: \\
(1) System interaction relies on the Streamlit framework, and the interaction experience in complex scenarios and the stability of large-scale multi-agent collaboration still need verification. \\
(2) Viewpoint evaluation and sentiment analysis heavily depend on specific LLMs, and the generalization ability and cross-domain applicability of the models need further validation.

\section{Acknowledgement}
We would like to thank all the participants who took part in our surveys and interviews.

\bibliographystyle{plainnat} 
\bibliography{reference} 


\appendix

\section{Survey design}

\subsection*{X System Usability Scale (SUS)}
This questionnaire aims to evaluate the performance of the X system in human-computer interaction design using the SUS scale.  
You can access the relevant functions of X through the following link: xxx  
Please select the \texttt{human on} section in the left sidebar and fill out this questionnaire after use.  
Pay attention to the performance of the intent alignment module, the supervised LLM expert discussion session, and the \texttt{cothinker} module during use.  

The benefits of large artificial intelligence models to social production and life are gradually emerging. Multi-agent systems and workflows built on LLMs have demonstrated a high degree of automation. However, in high-risk fields such as medical and financial decision-making scenarios, human participation in decision-making and aligning intelligent systems with human intent remain crucial.  
This work focuses on financial scenarios and builds a human-centered multi-agent financial analysis workflow based on Streamlit. It achieves task planning according to user intent through a BDI theory-based human-computer interaction method and a human-centered multi-agent discussion framework. The system also reduces the cognitive load of human users for complex systems through real-time updated structured text summaries and interactive thinking assistants. The workflow design reasonably integrates general large models and reasoning large models to enhance its ability to handle complex problems.

\subsubsection*{Questionnaire Items}
1. I think I would like to use this system frequently [Single choice] *  
   Strongly disagree \quad\textcircled{1}\quad\textcircled{2}\quad\textcircled{3}\quad\textcircled{4}\quad\textcircled{5}\quad Strongly agree  

2. I think this system is too complex [Single choice] *  
   Strongly disagree \quad\textcircled{1}\quad\textcircled{2}\quad\textcircled{3}\quad\textcircled{4}\quad\textcircled{5}\quad Strongly agree  

3. I find this system easy to use [Single choice] *  
   Strongly disagree \quad\textcircled{1}\quad\textcircled{2}\quad\textcircled{3}\quad\textcircled{4}\quad\textcircled{5}\quad Strongly agree  

4. I need professional help to use this system [Single choice] *  
   Strongly disagree \quad\textcircled{1}\quad\textcircled{2}\quad\textcircled{3}\quad\textcircled{4}\quad\textcircled{5}\quad Strongly agree  

5. I think the different functions of this system are well integrated [Single choice] *  
   Strongly disagree \quad\textcircled{1}\quad\textcircled{2}\quad\textcircled{3}\quad\textcircled{4}\quad\textcircled{5}\quad Strongly agree  

6. I think this system is too inconsistent [Single choice] *  
   Strongly disagree \quad\textcircled{1}\quad\textcircled{2}\quad\textcircled{3}\quad\textcircled{4}\quad\textcircled{5}\quad Strongly agree  

7. I think most people will learn to use this system quickly [Single choice] *  
   Strongly disagree \quad\textcircled{1}\quad\textcircled{2}\quad\textcircled{3}\quad\textcircled{4}\quad\textcircled{5}\quad Strongly agree  

8. I find this system very clumsy to use [Single choice] *  
   Strongly disagree \quad\textcircled{1}\quad\textcircled{2}\quad\textcircled{3}\quad\textcircled{4}\quad\textcircled{5}\quad Strongly agree  

9. I feel confident using this system [Single choice] *  
   Strongly disagree \quad\textcircled{1}\quad\textcircled{2}\quad\textcircled{3}\quad\textcircled{4}\quad\textcircled{5}\quad Strongly agree  

10. I need to learn a lot before using this system [Single choice] *  
    Strongly disagree \quad\textcircled{1}\quad\textcircled{2}\quad\textcircled{3}\quad\textcircled{4}\quad\textcircled{5}\quad Strongly agree  

11. Can the intent alignment module recognize and materialize your relevant intent? [Single choice]  
    \textcircled{Y}es \quad \textcircled{N}o \quad \textcircled{O}pinion  

12. Does interacting with the Cothinker module during supervised LLM expert discussions bring you inspiration? [Single choice]  
    \textcircled{Y}es \quad \textcircled{N}o \quad \textcircled{O}pinion  

13. Does structured text (Mindmap) reduce your cognitive load during supervised LLM expert discussions? [Single choice]  
    \textcircled{Y}es \quad \textcircled{N}o \quad \textcircled{O}pinion  

14. Did supervised LLM expert discussions reveal more novel perspectives on problem analysis? [Single choice]  
    \textcircled{Y}es \quad \textcircled{N}o \quad \textcircled{O}pinion

15. If you have any questions, please fill them here or contact us  
    Email: xxx [Fill-in-the-blank]  
    \underline{\makebox[8cm]{}}  

16. You may leave your name or nickname (may appear in acknowledgments if applicable) [Fill-in-the-blank]  
    \underline{\makebox[8cm]{}}

\subsection*{X System NASA-TLX Evaluation Scale}
This questionnaire aims to evaluate the performance of the X system in human-computer interaction design using the SUS scale.  
You can access the relevant functions of X through the following link: xxx  
Please select the \texttt{human on} section in the left sidebar and fill out this questionnaire after use.  
Pay attention to the performance of the intent alignment module, the supervised LLM expert discussion session, and the \texttt{cothinker} module during use.  

The benefits of large artificial intelligence models to social production and life are gradually emerging. Multi-agent systems and workflows built on LLMs have demonstrated a high degree of automation. However, in high-risk fields such as medical and financial decision-making scenarios, human participation in decision-making and aligning intelligent systems with human intent remain crucial.  
This work focuses on financial scenarios and builds a human-centered multi-agent financial analysis workflow based on Streamlit. It achieves task planning according to user intent through a BDI theory-based human-computer interaction method and a human-centered multi-agent discussion framework. The system also reduces the cognitive load of human users for complex systems through real-time updated structured text summaries and interactive thinking assistants. The workflow design reasonably integrates general large models and reasoning large models to enhance its ability to handle complex problems.

\subsubsection*{Evaluation Dimensions and Questions}
\textbf{Section 1: Core Evaluation Questions}  
1. \textbf{Mental Demand}  
   How much mental and perceptual effort did you exert while using X (e.g., thinking, decision-making, calculating, remembering, searching)? Was the task mentally easy or difficult, simple or complex for you? [Single choice] *  

2. \textbf{Physical Demand}  
   How much physical effort did you exert while using X (e.g., pushing, pulling, turning, controlling)? Was the task physically easy or difficult, light or strenuous for you? [Single choice] *  

3. \textbf{Temporal Demand}  
   How much time pressure did you feel due to the task pace while using X? Was the task pace slow or fast? [Single choice] *  

4. \textbf{Effort Level}  
   How much effort (mental and physical) did you expend to complete the task while using X? [Single choice] *  

5. \textbf{Overall Performance}  
   Were you satisfied with X's performance during use? [Single choice] *  

6. \textbf{Frustration Level}  
   How much frustration, stress, or irritation did you feel while using X? [Single choice] *  

\textbf{Additional Information}  
- If you have any questions, please fill them here or contact us  
  Email: xxx [Fill-in-the-blank]  
  \underline{\makebox[8cm]{}}  

- You may leave your name or nickname (may appear in acknowledgments if applicable) [Fill-in-the-blank]  
  \underline{\makebox[8cm]{}}  

\section{Interview design}
\section*{Interview Design for X System Usability Evaluation}

\subsection*{Interview Objectives}
This interview aims to assess the X system’s performance in human-computer interaction design, focusing on dimensions such as intuitiveness, intent alignment, learning curve, emotional experience, reliability, and personalization. The interview is expected to last 30–45 minutes, with content used solely for research purposes and all data strictly confidential with no associated risks.

\subsection*{Interview Outline}

\subsubsection*{1. Intuitiveness of Human-Computer Interaction}
- When using the system, were you able to quickly understand the system's feedback and prompts? Please describe your specific experience.  
- Did the system's interface design make the operation flow feel natural and intuitive? Please provide examples.  
- Did unclear interface design cause any operational difficulties? If so, describe the specific situation.  

\subsubsection*{2. Intent Alignment and System Feedback}
- Do you feel the system's feedback aligned with your operational intentions? If there were inconsistencies, describe the specific scenarios.  
- Could the system accurately predict your needs and provide proactive support? Please give examples.  
- During operation, did the system misunderstand your intentions? Describe the specific scenarios where this occurred.  

\subsubsection*{3. System Learning Curve}
- How difficult was the system to learn? Was additional training required to use it proficiently?  
- When using the system, did you feel your operational efficiency gradually improving? Describe your experience.  
- Were you able to quickly master the system's new features? If there were difficulties, explain the specific reasons.  

\subsubsection*{4. System Reliability and Trust}
- Do you trust the system's decisions and feedback? Please explain the reasons.  
- When the system made errors, were you able to understand and correct them? Describe your experience.  
- Do you consider the system stable enough for daily use? If there were instability issues, provide examples.  

\subsubsection*{5. Emotional Experience and User Satisfaction}
- Did using the system evoke feelings of pleasure or frustration? Describe specific scenarios.  
- What is your overall satisfaction with the system? Do you have any improvement suggestions?  
- Did the system surprise or disappoint you in any aspects? Please provide examples.  

\subsubsection*{6. Personalization and Adaptability}
- Do you feel the system can adapt to your personal needs? Please provide examples.  
- Can the system accommodate usage requirements in different scenarios? Describe your experience.  
- Would you like the system to offer more personalized features? If so, specify your needs.  

\subsubsection*{7. Additional Suggestions and Comments}
- What other comments or suggestions do you have regarding the overall user experience of the system?  
- If you were to design an ideal human-computer interaction system, how would you improve the current system?  
- Is there any other experience or thought you would like to share?

\subsection*{Interview Guidelines}
1. During the interview, interviewers should maintain a neutral attitude and avoid asking leading questions.  
2. Interviewers should record interviewees’ non-verbal information (e.g., expressions, tone) to supplement the interview content.  
3. After the interview, thank participants and briefly explain the follow-up research plan.

\subsection*{Core Dimensions of the Interview Framework}
1. **System Intuitiveness**  
   - Evaluation Focus: Interface understandability, operational naturalness  
   - Example Question: "Were the system's feedback prompts clear? Describe your specific experience."  

2. **Intent Alignment**  
   - Evaluation Focus: Consistency between operations and feedback  
   - Example Question: "Could the system accurately predict your needs? Please provide examples."  

3. **Emotional Experience**  
   - Evaluation Focus: Pleasure or frustration during use  
   - Example Question: "Which designs surprised or disappointed you?"  

\section{Interview sentiment design based on LLMs}
We conducted research on the quantitative analysis of sentiment in interview data and proposed a method based on LLM next token prediction. We selected five large language models for sentiment analysis: \textbf{Qwen2.5-7B-Guba-Senti}, \textbf{Qwen2.5-7B-Instruct}, \textbf{Qwen2.5-14B-Instruct}, \textbf{DeepSeek-R1-Distill-Qwen-7B}, and \textbf{DeepSeek-R1-Distill-Qwen-14B}. Among these, Qwen2.5-7B-Guba-Senti is a vertically fine-tuned model that achieved 85\% three-class accuracy on the test set for sentiment classification of Eastmoney stock bar commentary data through supervised fine-tuning of the Qwen2.5-7B model. This significantly outperformed FinBERT and other large models, effectively improving the performance of sentiment analysis in financial social media. The DeepSeek-R1-Distill-Qwen models are lightweight models derived by DeepSeek company through knowledge distillation of their R1 model using Qwen as the base model, exhibiting a certain degree of reasoning ability. However, according to the scaling law, the capabilities of LLMs are generally positively correlated with their parameters.

All the LLM models we used are based on the Qwen base model. By querying their vocabulary, we found that the positive and negative tokens exist in this series of models. We guided the LLMs to model the semantics of the interview content through prompting. The prompt used is as follows. This approach leverages the autoregressive nature of LLMs in predicting the next token. Following natural language logic, after inputting the prompt, the LLM should predict either the positive or negative token. This effectively models and represents the sentiment of the interview content using the LLM, and the sentiment score is then quantified by analyzing the probability of these next tokens during decoding.

\subsection{Prompt Design for Different Models}

\subsubsection{General Model Prompt Design}
  \texttt{input\_text = "..." } \\
  \texttt{\{data[key]\} } \\
  \texttt{what emotion does this passage express?} \\
  \texttt{(Please choose one from 'positive' and 'negative' only)} \\
  \texttt{Answer: } \\
  \texttt{...} \\

\subsubsection{Reasoning Model Prompt Design}
  \texttt{input\_text = "..." } \\
  \texttt{\{data[key]\} } \\
  \texttt{what emotion does this passage express?} \\
  \texttt{(Please choose one from 'positive' and 'negative' only; do not generate any other content)} \\
  \texttt{</think>} \\
  \texttt{...} \\

In the experiment, the top 10 candidate next tokens generated by the LLM were extracted. After retaining the probabilities $P=[p_1, p_2, \ldots, p_{10}]$ of the top 10 tokens after the softmax layer, the word vectors $V_i$ of these top 10 tokens were extracted from the LLM (the top 10 tokens generally include the two tokens "positive" and "negative"). However, from the perspective of similarity analysis, the tokens "positive" and "negative" are mutually exclusive. Therefore, extracting the top 10 tokens for similarity calculation is actually a way to avoid the obtained quantitative analysis results being completely one-sided. The cosine similarity between the word vectors and the tokens "positive" $V_{pos}$ and "negative" $V_{neg}$ were calculated separately to obtain the similarity between the top 10 tokens and positive and negative, denoted as $C \equiv [C_{pos}, C_{neg}]^T$, where

\begin{align}
C_{pos} &= [\text{similarity}(V_{pos}, V_1), \ldots, \text{similarity}(V_{pos}, V_{10})] \\
C_{neg} &= [\text{similarity}(V_{neg}, V_1), \ldots, \text{similarity}(V_{neg}, V_{10})]
\end{align}

\begin{figure}[ht] 
  \centering 
  \includegraphics[width=0.95\linewidth]{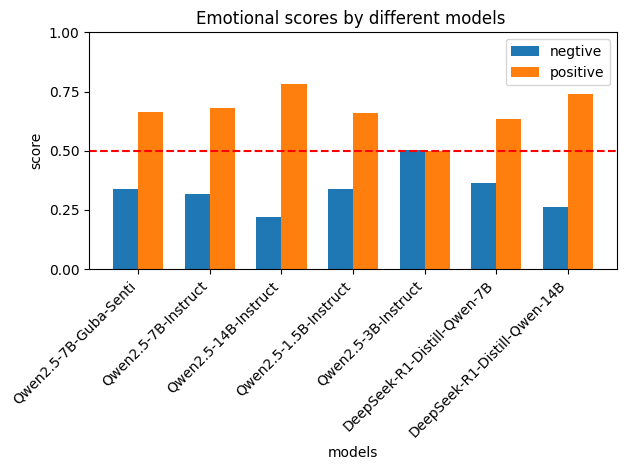}
  \caption{Emotional scores} 
  \label{fig:Emotional scores} 
\end{figure}

The final score is:

\begin{equation}
S = P^T C
\end{equation}

where $S = [s_{pos}, s_{neg}]$, $s_{pos}$ represents the positive sentiment score, and $s_{neg}$ represents the negative sentiment score. Finally, the two are normalized. During the quantitative calculation of sentiment, in order to eliminate the bias that a single model may produce, the calculation results of multiple models were averaged. Figure \ref{fig:Emotional scores} shows the sentiment scores calculated by different models.

\end{document}